# Strong Decays of Charmonia

Zaki Ahmad

*Centre For High Energy Physics, Punjab University, Lahore(54590), Pakistan.* [*]

Ishrat Asghar

*University of Education, Lahore, Faisalabad Campus, Faisalabad*[†]

Faisal Akram[‡] and Bilal Masud[§]

*Centre For High Energy Physics, Punjab University, Lahore(54590), Pakistan.*

(Dated: April 7, 2025)

## Abstract

In this work, we calculate the charmonium spectrum and strong decay widths of $c\bar{c}$ states. The calculations are performed using the quark potential model with which incorporates the relativistic effects. The resulting $c\bar{c}$ spectrum exhibits a good agreement with experimental data. The open flavor strong decay widths are calculated employing the $^3P_0$ model. We adopt two choices of wave functions to determine the observables: the first involves the utilization of realistic wave functions obtained by solving relativistic Schrödinger equation, while the second employs Simple harmonic oscillator (SHO) wave functions whose parameters are fitted to the realistic wave functions. We find that the decay widths of higher charmonia states above the threshold value of open charm mesons are well described with these wave functions. We provide a comprehensive analysis of the strong decay widths and assigned specific charmonium states: $X(3940)$ is assigned to the $\eta_c(3S)$ state, $Y(4660)$ to the $\psi(5S)$ state, $X(3915)$ to the $\chi_0(2P)$ state, $Z_c(3900)$ to the $h_c(2P)$ state, $X(4350)$ to the $\chi_2(3P)$ state, and $X(3842)$ to the $\psi_3(1D)$ state. We also compare our results with available experimental data and theoretical predictions of other models.

---

[*] zakiahmad754@gmail.com
[†] ishrat.asghar@ue.edu.pk
[‡] faisal.chep@pu.edu.pk
[§] bilalmasud.chep@pu.edu.pk

# I. INTRODUCTION

The $J/\psi$ meson, discovered in 1974 independently at the Stanford Linear Accelerator Center (SLAC) [1] and Brookhaven National Laboratory (BNL) [2], marks a pivotal moment in the history of particle physics. The discovery had a profound implications for quantum chromodynamcis, particularly by supporting the notion of confinement and opening the door to the study of heavy quark dynamics. Subsequent discoveries showed that the bound states of $c\bar{c}$ can be categorized by spectroscopic notations $n^{2S+1}L_J$, where $n$ is the principle quantum number corresponding to radial excitation, $S$ is spin of quark anti-quark pair, and $L$ is the relative orbital angular momentum. The existence of these states and their corresponding $J^{PC}$ are well described by the quark model assuming a pair of $c\bar{c}$ dynamical quarks. However, the recent discoveries of the X, Y, Z states have brought a new twist to our understanding of charmonium spectrum. These exotic states, primarily discovered by LHCb, Belle, BaBar, and BESIII, do not conform to the conventional quark model, whilst allowed in the realm of quantum chromodynamics. They exhibit properties that suggest they could be tetraquarks, molecules of mesons, hybrid states or other exotic configurations. Understanding the nature of these states has opened a new challenge in hadron spectroscopy.

In this work, we study the mass spectra and open flavor strong decays of charmonium using a quark anti-quark potential model that incorporates relativistic effects [4]. We compare our predicted masses with available experimental data from [3] and theoretical results from [4]. Since the final states of charmonium strong decays are open charm mesons $D$ and $D_s$, we solve relativistic Schrödinger equation also for $c\bar{q}$ ($q = u, d$) and $c\bar{s}$ systems using the same relativistic potential adjusted for unequal quark masses. To calculate the widths of charmonium strong decays, we employ the $^3P_0$ decay model. This approach enables us to identify most of the conventional charmonium states by comparing our predictions with experimental data.

A detailed study of strong decays of light mesons using the $^3P_0$ model was presented in reference [5]. These work explored a wide range of strong decay amplitudes, assuming SHO wave functions. This approach was later extended to other sectors, such as $q\bar{s}$, $s\bar{s}$, $b\bar{b}$, $c\bar{c}$, $c\bar{q}$, $b\bar{q}$, and $b\bar{s}$ (where $q = u, d$) [6–10]. In Ref. [11] strong decay widths are calculated by using the numerical wave functions for the initial meson and SHO wave functions for the final states. The numerical wave functions in this work are calculated by using linear potential model (LP) without incorporating relativistic effects. Given the approximation in wave functions, the next significant improvement is to incorporate numerically calculated relativistic wave functions of initial and final states in calculating strong decay widths. This we do in the present work. To study the impact of using these numerical wave functions, we compare our results with those obtained through using SHO wave functions and other theoretical results [7, 11].

The paper is organized as follow: In section. II, we describe spectroscopy in detail. First, the quark anti-quark potential model is explained to calculate the meson masses. Wave equation of quark anti-quark and its solution is described in section II B. In section. III we review the strong decay model to calculate the strong decay widths by using the realistic and SHO wave functions. In section. IV we have reported the strong decay widths and available experimental data. Finally, we have concluded our work in section V.

# II. SPECTROSCOPY

## A. Potential model for quark anti-quark system

In this section, we describe the potential model that we use to study the quark anti-quark system. It is a variant of the one used in the Ref. [4]. In the rest frame of the $q\bar{q}$ system, relativistic type Schrödinger wave equation is given as

$$H\psi(\mathbf{r}) = E\psi(\mathbf{r}). \tag{1}$$

The effective Hamiltonian $H$ of the system is

$$H = \sqrt{p_1^2 + m_1^2} + \sqrt{p_2^2 + m_2^2} + V_{q\bar{q}}(r), \tag{2}$$

where $m_1$, $p_1$ are the mass and momentum of the quark and $m_2$, $p_2$ are the mass and momentum of the anti-quark. The central potential $V_{q\bar{q}}(r)$ is given as

$$V_{q\bar{q}}(r) = V_0 + V_{\text{hyp}} + V_{\text{so}} + V_{\text{tensor}}, \tag{3}$$

where $V_0$ is the spin independent part which is given as [4, 12]:

$$V_0 = -\frac{4}{3}\frac{\alpha_s(r)}{r} + br + c. \tag{4}$$

The first term stems from the one gluon exchange diagram of pQCD and second one is the linear confinement term with the string tension $b$, and $c$ is a phenomenological constant, which can be adjusted to give the correct ground state energy level of the $q\bar{q}$ system. In equation (4), the running coupling constant $\alpha_s(r)$ is the Fourier transform of $\alpha_s(Q^2)$. A convenient parametrization of $\alpha_s(r)$ is given as [4]

$$\alpha_s(r) = \sum_i \alpha_i \frac{2}{\sqrt{\pi}} \int_0^{\gamma_i r} e^{-x^2} dx. \tag{5}$$

In this work we take $2\gamma_1 = 1$, $2\gamma_2 = \sqrt{10}$, and $2\gamma_3 = \sqrt{1000}$ [4]. For this choice the parameters, $\alpha_1$ and $\alpha_2$ control the shape of $\alpha_s(r)$ at short distance, where it is known by pQCD. Whereas $\alpha_3$ control its shape at large distance, where QCD is non-perturbative. so, we leave $\alpha_3$ as a free parameter. Equivalently, we can define the parameter $\alpha_s = \sum_i \alpha_i$, and treat it as a free parameter in exchange of $\alpha_3$. The second term in equation (3) is the spin hyperfine interaction given as:

$$V_{\text{hyp}} = \frac{32\pi}{9 m_1 m_2} \alpha_s(r) \delta_a(r) \vec{s}_1 . \vec{s}_2, \tag{6}$$

with

$$\delta_\sigma(r) = \left(\frac{\sigma_h}{\sqrt{\pi}}\right)^3 e^{-\sigma_h^2 r^2},$$

where $\sigma_h$ is a phenomological constant. The spin orbit term in the potential is written as:

$$V_{so} = \frac{4}{3} \frac{\alpha_s(r)}{r} \left[ \left(\frac{1}{2m_1^2} + \frac{1}{m_1 m_2}\right) \vec{L}.\vec{s}_1 + \left(\frac{1}{2m_2} + \frac{1}{m_1 m_2}\right) \vec{L}.\vec{s}_2 \right] - \frac{b}{r} \left( \frac{L.s_1}{2m_1^2} + \frac{L.s_2}{2m_2} \right), \tag{7}$$

where $\mathbf{L}$ is the relative orbital angular momentum between quark anti-quark and the tensor term is

$$V_{\text{tensor}} = \frac{\alpha_s(r)}{m_1 m_2} \frac{1}{r^3} \left( \vec{s}_1.\hat{r} \vec{s}_2.\hat{r} - \frac{1}{3} \vec{s}_1.\vec{s}_2 \right). \tag{8}$$

In strong decay of charmonium mesons, the final states contain open charm mesons. The effective $q\bar{q}$ potential for these states is given as [12]

$$V_{q\bar{q}}(r) = -\frac{4}{3} \frac{\alpha_s(r)}{r} + br + c + \frac{32\pi}{9} \alpha_s(r) \frac{\mathbf{s}_1.\mathbf{s}_2}{m_1 \tilde{m}_{2a}} \left(\frac{\sigma_h}{\pi^{1/2}}\right)^3 e^{-\sigma_h^2 r^2} + \frac{4\alpha_s}{3r^3}$$
$$\left[ \left(\frac{1}{2m_1^2} + \frac{1}{m_1 \tilde{m}_{2c}}\right) \mathbf{s}_1.\mathbf{L} + \left(\frac{1}{2\tilde{m}_{2c}} + \frac{1}{m_1 \tilde{m}_{2c}}\right) \mathbf{s}_2.\mathbf{L} \right] - \frac{b}{r} \left( \frac{\mathbf{s}_1.\mathbf{L}}{2m_1^2} + \frac{\mathbf{s}_2.\mathbf{L}}{2\tilde{m}_{2d}} \right) + \frac{\alpha_s(r)}{m_1 \tilde{m}_{2b}} \frac{1}{r^3} \left( \vec{s}_1.\hat{r}\vec{s}_2.\hat{r} - \frac{1}{3}\vec{s}_1.\vec{s}_2 \right). \tag{9}$$

Where we have used the light quark mass $m_2$ by $\tilde{m}_{2i}$, $i = a,b,c,d$ to incorporate relativistic corrections in the heavy light system [12]. It is to be noted that potential is strongly attractive at short distance and the resultant wave function becomes unstable. To overcome this problem, we apply the smearing in position coordinates. The Corresponding smeared function has following form [4]

$$\rho(r) = \int d^3 r' \frac{\sigma_s^3}{\pi^{3/2}} e^{-\sigma_s^2 (r'-r)^2}, \tag{10}$$

where $\sigma_s$ is an another phenomenological parameter. We determine the parameters of the model by fitting to experimental masses of $c\bar{c}$ and $c\bar{q}$, $q = u,d,s$. The numerical values of these parameters are discussed in section (II D).

### B. Solution of the wave equation

In Ref. [12] solutions of the wave equation are calculated without incorporating the quark spin dependent interactions. The effect of these interactions is incorporated in the meson masses as a perturbative correction. In this paper, we solve the equation (1) including the spin dependent interactions without resorting to any perturbative method. Consequently the effect of spin dependent interactions appears not only in the mass but also in the corresponding wave function. Like Ref. [12], we re-express, Eq. (1) as following

$$\frac{2}{\pi}\int dk k^2 \int dr' r r' \left(\sqrt{k_1^2 + m_1^2} + \sqrt{k_2^2 + m_2^2}\right) j_l(kr) j_l(kr') R_l(r') + V_{q\bar{q}}(r) R_l(r) = E R_l(r). \tag{11}$$

In equation(11), $R_l(r)$ is the radial part of the wave function and stated as:

$$R_l(r) = \sum_{n=1}^{\infty} c_n \frac{a_n r}{L} j_l\left(\frac{a_n r}{L}\right), \tag{12}$$

where $c'_n s$ are the expansion coefficients and $a_n$ is the $n$th root of the spherical bessel function $j_l(a_n) = 0$. In above equation, summation is to be truncated at a large integer $N$ in the numerical calculation and momentum is to discritized over $k$ and replaced by a summation. Equation (11) has becomes

$$\frac{2}{\pi L^3}\nabla a_n a^2 n N_n^2 \left(\sqrt{\left(\frac{a_n}{L}\right)^2 + m_1^2} + \sqrt{\left(\frac{a_n}{L}\right)^2 + m_2^2}\right) c_n + \sum_{n=1}^{N} \frac{a_m}{N_n^2 a_n} \int_0^L dr V_{q\bar{q}}(r) r^2 j_l\left(\frac{a_n r}{L}\right) j_l\left(\frac{a_m r}{L}\right) c_m = E c_n, \tag{13}$$

where $\nabla a_n = a_n - a_{n-1}$ and $N_m^2$ represents the module of spherical bessel function $j_l$ and is given as:

$$N_m^2 = \int_0^L dr' r'^2 j_l\left(\frac{a_m r'}{L}\right). \tag{14}$$

Equation (13) is the eigen equation of the hamiltonian $H$ in the matrix form. The solution of equation (13) are depends on the value of $N$ and $L$ and we have chosen the $N \geq 50$ and $L \geq 5$ fm in the numerical calculation. Here we have $N$ basis of the $|^{2S+1}L_J>$ for the heavy quark limit. $|^{2S+1}L_J>$ is the eigenvector of equation (1), where $J$ is the total general angular momentum of meson, $S$ is the total spin of quark anti-quark and $l$ is the orbital angular momentum between the quark and the anti-quark. The coefficients obtained from equation (13) used in equation (12) to obtain the numerical wave function in position space. In heavy quark limit the orbital angular momentum $L$ and spin angular momentum $S$ have definite value. So, equation (13) is used to find the eigenfunctions and eigenvalues of the heavy quark system. In heavy light system the charge parity is violate due to the light quark and meson state has not definite value of orbital angular momentum $L$ and spin angular momentum $S$. So, mixing occurs between these states. In heavy light system $V_{tensor}$ in equation (8) does not conserve the orbital angular momentum and it causes the mixing of the orbital angular momenta between the states. For tensor mixed states, the wave function $\psi(r)$ of radial and angular part has following representation

$$\psi(r) = \sum_n c_n^{(0)} \frac{a_n}{L} j_l\left(\frac{a_n r}{L}\right) |^3 L_J > + \sum_n c_n^{(1)} \frac{a_n}{L} j_{l+2}\left(\frac{a_n r}{L}\right) |^3 L'_J >, \tag{15}$$

where $c_n^{(0)}$ and $c_n^{(1)}$ are expansion coefficients associated with $|^3 L_J>$ and $|^3 L'_J>$ states which are given as

$$|^3 L_J> = |1, 1>_s Y_{ll}(\hat{r}), \tag{16}$$

and

$$|^3 L'_J> = \sum_{m_s=-1,0,1} C_{m_s} |1, m_s> Y_{l+2, l-1-m_s}(\hat{r}). \tag{17}$$

Substituting equation (15) into equation (1), we find the two coupled equations for the tensor mixed states. The resultant expressions are

$$\frac{2}{\pi L^3}\nabla a_n a_n^2 N_n^2 \left(\sqrt{\left(\frac{a_n}{L}\right)^2 + m_1^2} + \sqrt{\left(\frac{a_n}{L}\right)^2 + m_2^2}\right) c_n^{(0)} + \sum_{n=1}^{N} \frac{a_m}{N_n^2 a_n}$$
$$\left(\int_0^L dr r^2 <V_{q\bar{q}}>_{11}(r) j_l\left(\frac{a_n r}{L}\right) j_l\left(\frac{a_m r}{L}\right) c_m^{(0)} + \int_0^L dr r^2 <V_{q\bar{q}}>_{12}(r) j_{l+2}\left(\frac{a_n r}{L}\right) j_l\left(\frac{a_m r}{L}\right) c_m^{(1)} + \right) = E c_n^{(0)}, \tag{18}$$

$$\frac{2}{\pi L^3}\nabla a_n a_n^2 N_n^2 \left(\sqrt{\left(\frac{a_n}{L}\right)^2 + m_1^2} + \sqrt{\left(\frac{a_n}{L}\right)^2 + m_2^2}\right) c_n^{(1)} + \sum_{n=1}^{N} \frac{a_m}{N_n^2 a_n}$$
$$\left(\int_0^L dr r^2 <V_{q\bar{q}}>_{21}(r) j_{l+2}\left(\frac{a_n r}{L}\right) j_l\left(\frac{a_m r}{L}\right) c_m^{(0)} + \int_0^L dr r^2 <V_{q\bar{q}}>_{22}(r) j_{l+2}\left(\frac{a_n r}{L}\right) j_{l+2}\left(\frac{a_m r}{L}\right) c_m^{(1)}\right) = E c_n^{(1)}. \tag{19}$$

In equation (18) and (19), $<V_{q\bar{q}}>_{ij}$, $i,j = 1,2$. are the potential energy matrix elements which are defined in the following form

$$<V_{q\bar{q}}(r)>_{11} = <^3L_J | V_{q\bar{q}}(r) |^3 L_J>, \tag{20}$$

$$<V_{q\bar{q}}(r)>_{12} = <^3L_J | V_{q\bar{q}}(r) |^3 L'_J>, \tag{21}$$

$$<V_{q\bar{q}}(r)>_{21} = <^3L'_J | V_{q\bar{q}}(r) |^3 L_J>, \tag{22}$$

$$<V_{q\bar{q}}(r)>_{22} = <^3L'_J | V_{q\bar{q}}(r) |^3 L'_J>. \tag{23}$$

The spin interaction term in equation (7) does not conserve the total spin of the quark anti-quark and it causes the mixing of the spin quantum number between different states. The wave function $\psi(r)$ for the spin mixed states has following form

$$\psi(r) = \sum_n c_n^{(0)} \frac{a_n}{L} j_l\left(\frac{a_n r}{L}\right) |^1 L_J> + \sum_n c_n^{(1)} \frac{a_n}{L} j_l\left(\frac{a_n r}{L}\right) |^3 L_J>, \tag{24}$$

where $|^1 L_J>$ and $|^3 L_J>$ states are defined as

$$|^1 L_J> = |0,0>_s Y_{ll}(\hat{r}), \tag{25}$$

and

$$|^3 L_J> = \sum_{m_s=0,1} C_{m_s} |1, m_s> Y_{l,l-m_s}(\hat{r}). \tag{26}$$

Substitute equation (24) into equation (1), we get the resultant expression for spin mixed states as

$$\frac{2}{\pi L^3} \nabla a_n a_n^2 N_n^2 \left(\sqrt{\left(\frac{a_n}{L}\right)^2 + m_1^2} + \sqrt{\left(\frac{a_n}{L}\right)^2 + m_2^2}\right) c_n^{(0)} + \sum_{n=1}^{N} \frac{a_m}{N_n^2 a_n}$$
$$\int_0^L dr\, r^2 \left(<V_{q\bar{q}}>_{11}(r) c_m^{(0)} + <V_{q\bar{q}}>_{12}(r) c_m^{(1)}\right) j_l\left(\frac{a_n r}{L}\right) j_l\left(\frac{a_m r}{L}\right) = E c_n^{(0)}. \tag{27}$$

$$\frac{2}{\pi L^3} \nabla a_n a_n^2 N_n^2 \left(\sqrt{\left(\frac{a_n}{L}\right)^2 + m_1^2} + \sqrt{\left(\frac{a_n}{L}\right)^2 + m_2^2}\right) c_n^{(1)} + \sum_{n=1}^{N} \frac{a_m}{N_n^2 a_n}$$
$$\int_0^L dr\, r^2 \left(<V_{q\bar{q}}>_{21}(r) c_m^{(0)} + <V_{q\bar{q}}>_{22}(r) c_m^{(1)}\right) j_l\left(\frac{a_n r}{L}\right) j_l\left(\frac{a_m r}{L}\right) = E c_n^{(1)}, \tag{28}$$

where the potential energy matrix elements in equation (27) and (28) are given as

$$<V_{q\bar{q}}(r)>_{11} = <^1L_J | V_{q\bar{q}}(r) |^1 L_J>, \tag{29}$$

$$<V_{q\bar{q}}(r)>_{12} = <^1L_J | V_{q\bar{q}}(r) |^3 L_J>, \tag{30}$$

$$<V_{q\bar{q}}(r)>_{21} = <^3L_J | V_{q\bar{q}}(r) |^1 L_J>, \tag{31}$$

$$<V_{q\bar{q}}(r)>_{22} = <^3L_J | V_{q\bar{q}}(r) |^3 L_J>. \tag{32}$$

### C. Hamiltonian ($H$) matrix of mixed states

The $H$ matrix of mixed states is described as

$$H = \begin{pmatrix} H_{11} & H_{12} \\ H_{21} & H_{22} \end{pmatrix}. \tag{33}$$

The order of $H$ matrix is $2N \times 2N$ where $H_{11}$, $H_{12}$, $H_{21}$ and $H_{22}$ are matrices of order $N \times N$. In case of tensor mixing, $H_{11}$ and $H_{12}$ are read from equation (18) while $H_{21}$ and $H_{22}$ are read from equation (19). In case of spin mixing, $H_{11}$ and $H_{12}$ are read from equation (27) while $H_{21}$ and $H_{22}$ are read from equation (28). The kinetic matrix elements contribute only in diagonal hamiltonian $H_{11}$ and $H_{22}$. Potential energy matrices $<V_{q\bar{q}}>_{11}$, $<V_{q\bar{q}}>_{12}$, $<V_{q\bar{q}}>_{21}$ and $<V_{q\bar{q}}>_{22}$ are terms of $H_{11}$, $H_{12}$, $H_{21}$ and $H_{22}$ respectively. It is to be noted that mixed states arise due to off diagonal hamiltonian matrices $H_{12}$ and $H_{21}$.

### D. Quark potential model parameter

The parameters for charmonium ($c\bar{c}$), $D$ and $D_s$ mesons sectors are reported in table (I) and (II). These values are obtained by fitting the available experimental states of mesons. In heavy light system the light quark mass $\widetilde{m}_{2j}$ are

TABLE I. Quark model parameters for charmonium fitted to the experimental available masses.

| Parameter | values |
|---|---|
| $m_c$ | 1.4905 GeV |
| $\alpha_s$ | 0.743 |
| $b$ | 0.1634 GeV$^2$ |
| $\sigma_h$ | 1.2364 GeV |
| $\sigma_s$ | 0.2409 GeV |
| $c$ | 0.0144 GeV |

depend on the quark anti-quark system. It can be written as:

$$\widetilde{m}_{2i} = \epsilon_i \widetilde{m}_2, \ i = a, b, c, d.$$

The fitting values of $\epsilon_i$ are

$$(\epsilon_1, \epsilon_2, \epsilon_3, \epsilon_4) = (1.0, 1.49, 1.49, 1.11) \tag{34}$$

and the values of $\widetilde{m}_2$ for $c\bar{q}$ and $c\bar{s}$ system are

$$\widetilde{m}_2 = \begin{cases} 0.427 \text{ GeV} & \text{for } c\bar{q} \text{ system} \\ 0.509 \text{ GeV} & \text{for } c\bar{s} \text{ system} \end{cases}. \tag{35}$$

The calculated mass spectrum of charmonium sector are reported in table (III) with the experimental known results

TABLE II. Quark model parameters for $D$ and $D_s$ sectors fitted to using the experimental available masses.

| Parameter | values |
|---|---|
| $m_u$ | 0.0312 GeV |
| $m_s$ | 0.3276 GeV |
| $\alpha_s$ | 0.4000 |
| $b$ | 0.1702 GeV$^2$ |
| $\sigma_h$ | 1.0652 GeV |
| $\sigma_s$ | 0.3852 GeV |
| $c$ | $-0.2348$ GeV |

[3] and spectrum of $D$ and $D_s$ sectors that are used in the strong decay of charmonium are reported in table (IV) and (V). We compare our predicted mass spectrum with the available experimental data and also with a relativized conventional meson model [4].

## III. THE DECAY MODEL OF STRONG DECAY

The Okubo-Zweig-Iizuka (OZI) allowed strong decay widths for the charmonium states above $D\overline{D}$ threshold will be determined by using the $^3P_0$ quark pair creation model which was firstly proposed by Micu in 1969 [13]. Later Le Yaouanc et al and collaborators [14–16] widely developed the model by applying it on different strong decays. The model has been extensively applied to strong decays of the charmonium states [7, 11, 17–19] etc. The basic assumption of the model is that a quark anti-quark pair is produced with vaccum quantum numbers ($0^{++}$). In the non-relativistic limit the interaction for the $^3P_0$ model is expressed as:

$$H_I = 2m\gamma \int \overline{\psi}\psi d^3x, \tag{36}$$

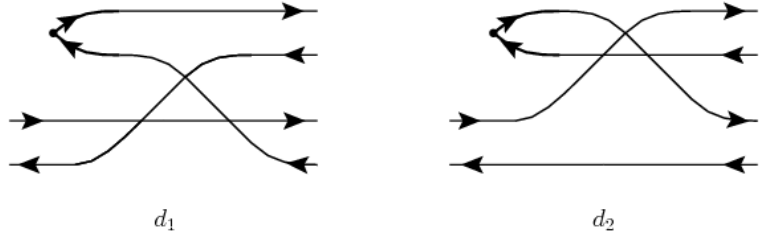

FIG. 1. Decay diagrams in $^3P_0$ Model.

where $m$ is the mass of the produced quark, $\gamma$ is the dimensionless free $^3P_0$ model parameter fitted to strong decay data $\psi$ represents the Dirac quark field which is given as

$$\psi(x) = \int \frac{d^3k}{(2\pi)^3} \left[ u(\vec{k},s) b(k) + v(-\vec{k},s) d^\dagger(-\vec{k}) \right] e^{i\vec{k}.x}. \tag{37}$$

To evaluate the decay width of a process $A \to BC$, we determine the decay amplitude $\langle BC|H_I|A \rangle$, where $|A\rangle$, $|B\rangle$ and $|C\rangle$ are the mesonic states of the quark model:

$$|A\rangle = |\mathbf{A}; nJM[LS]; II_z\rangle = \int d^3a\, d^3\overline{a}\, \delta(\mathbf{A} - \mathbf{a} - \overline{\mathbf{a}})\, \phi_{nL}\left(\frac{m_{\overline{a}}\mathbf{a} - m_a \overline{\mathbf{a}}}{m_a + m_{\overline{a}}}\right) \tag{38}$$
$$X^{JM[LS]}_{c,s,f;\overline{c},\overline{s},\overline{f}}\, Y_{LM_L}(\hat{k})\, b^\dagger_{c,s,f}(\mathbf{a})\, d^\dagger_{\overline{c},\overline{s},\overline{f}}(\overline{\mathbf{a}})\, |0\rangle,$$

where $\phi$ is the wave function that depends on the momenta $a$ and $\overline{a}$ of the quark and anti-quark with masses $m_a$ and $m_{\overline{a}}$. $X_{c,s,f;\overline{c},\overline{s},\overline{f}}$ matrix is given by:

$$X^{JM[LS]}_{c,s,f;\overline{c},\overline{s},\overline{f}} = \frac{\delta_{c\overline{c}}}{\sqrt{3}}\, \Xi^{I,I_z}_{f,\overline{f}}\, \langle \frac{1}{2}s, \frac{1}{2}\overline{s}|SM_S\rangle\, \langle SM_S, LM_L|JM\rangle.$$

A sum over repeated indices is understood in Eq.(38). In the last equation $\Xi^{I,I_z}_{f,\overline{f}}$ is a flavor wave function. In equation (36) the $b^\dagger d^\dagger$ leads the quark anti-quark production term in $^3P_0$ decay model. In each decay process are four diagrams contributes. But two of the diagrams has been suppressed by the OZI rule. The OZI allowed diagram are shown in figure 1. As in ref [20], if the produced quark goes into meson B, we call it diagram $d_1$ and if it goes to meson C, call it diagram $d_2$. The matrix element for each diagram factorises according to

$$\langle BC|H_I|A\rangle = I_{signature} I_{flavor} I_{spin+space}. \tag{39}$$

The $I_{signature}$ is, in the notation of [20]

$$I_{signature} = <0|\, b_c d_{\overline{c}} b_b d_{\overline{b}} b^\dagger_k d^\dagger_{-k} d^\dagger_{\overline{a}} b^\dagger_a\, |0>. \tag{40}$$

The flavor factors $I_{flavor}(d_1)$ and $I_{flavor}(d_2)$ are the overlap of the flavor wave functions of the initial meson and the system with created $(q\overline{q})$ pair for a specific charge channel. For a decay process having several charge channels for example $X_c \to D^+D^-$ and $X_c \to D^0\overline{D}^0$, we have to sum over all the charge channels. This can be described as a multiplication with a flavor multiplicity factor $\mathcal{F}$. The flavor factor $I_{flavor}(d_1)$ for each process discussed in this paper is zero, so we need to calculate the space and spin factors for diagram $(d_2)$ only. For all the processes discussed in this work, $I_{flavor}(d_2)$ and $\mathcal{F}$ are reported in Table(VI). Substituting equation (36)-(38) in $\langle BC|H_I|A\rangle$, we get the (remaining) spin and space factors of the diagram (d2) for the process $c\overline{c} \to c\overline{q} + q\overline{c}$, where $q = (u,d,s)$. In the centre of mass frame of meson-A ($|\mathbf{A}| = 0$ and $|\mathbf{B}| = |\mathbf{C}|$), the $I_{spin+space}(d_2)$ part of matrix element $\langle BC|H_I|A\rangle$ is given by

$$I_{spin+space}(d_2) = -2\gamma \int \frac{d^3\mathbf{P}}{(2\pi)^3}\, (P_z - B_z)\, \phi_A(\mathbf{P})\, \phi^*_B(\mathbf{P} - u\mathbf{B})\, \phi^*_C(\mathbf{P} - u\mathbf{B}) \frac{1}{m_q} <\overline{s}|\vec{\sigma}|\overline{s}>.\mathbf{k}, \tag{41}$$

where $\frac{1}{m_q} <\overline{s}|\vec{\sigma}|\overline{s}>.\mathbf{k}$ spin factor is to be calculated by using the equations (B14)- (B16) of ref. [20], and $u = \frac{m_c}{m_c + m_q}$, $(q \equiv u,d,s)$.

## A. $^3P_0$ model parameter fitting

In this work, the $^3P_0$ model parameter ($\gamma$) is determined by fitting it to the experimental decay widths of well established known charmonium states above the $D\bar{D}$ threshold i.e. $\psi(3\,^3S_1)$, $\psi(4\,^3S_1)$, $\psi(1\,^3D_1)$, $\psi(2\,^3D_1)$, $\chi_{c2}(2\,^3P_2)$. The unknown parameter $\gamma$ depends on choice of mesonic wave functions. we have used two different choices for the mesonic wave functions in our work:

1. Taking SHO wave functions of $\phi_A$, $\phi_B$ and $\phi_C$, as in [5, 20] in the form

$$\phi_{nlm}(\mathbf{k}) = (2\pi)^{3/2}(-i)^{2n+l}N_{nlm}k^l Y_{lm}(\hat{k})L_n^{l+1/2}(\beta^{-2}k^2)e^{-\frac{1}{2}\beta^{-2}k^2}, \qquad (42)$$

   where $\beta$ is the SHO parameter, $N_{nlm}$ is the normalization constant and $L_n^{l+1/2}$ is the associated Legendre polynomial. The open-flavor decay amplitudes of $c\bar{c}$ states were evaluated in the $^3P_0$ model by T. Barnes et al. [7]. We are using quite similar approach with following difference: 1)- We use quark potential model by including the relativistic effects for spectroscopy. 2)- We use SHO parameter $\beta_A$, $\beta_B$ and $\beta_C$ for meson $A$, $B$ and $C$ respectively rather than a universal value of $\beta$ parameter for all the mesons(initial+final). In this work, we obtain a $\beta$ parameter by fitting the SHO wave function to the quark model wave function calculated in section II. SHO fitted $\beta$ values of $c\bar{c}$, $D$ and $D_s$ mesons are reported in tables III, IV, V.

2. The SHO wave function is an approximation of the realistic wave function that provides the analytical results of our calculation in many cases; this lacks realistic results. In ref. [11] numerical wave function is used for initial states and for final states meson simple harmonic wave function with a universal value of SHO parameter $\beta$ is adopted to calculate the strong decay widths. We use the realistic wave functions of meson $A$, $B$ and $C$ obtained as a solution of the Schrödinger equation by using the quark potential model by incorporating the relativistic effects.

We obtain $\gamma = 0.323$ with realistic wave function and $\gamma = 0.337$ with SHO wave function after fitting to the available strong decay data. Our fitting parameter is close to the previous study of charmonium strong decays [7, 11] and reasonably described the decay widths properties. Strong decays widths of higher charmnomium states are listed in tables VIII, IX, X, XI, XII.

Finally, the decay amplitude can be combined with relativistic phase space to give the differential decay width, which is

$$\Gamma = 2\pi \frac{PE_B E_C}{M_A} \sum_{LS} |M_{LS}|^2, \qquad (43)$$

where $E_B$ and $E_C$ are the energies of final state mesons and $M_A$ is the rest mass of initial meson.

## IV. RESULTS AND DISCUSSIONS

Strong decay widths for the $3S$, $4S$, $5S$, $1P$, $2P$, $3P$, $1D$, $2D$, $1F$ are given in tables (VIII)-(XII). First we will discuss the well known experimental $c\bar{c}$ states above the threshold (3.73) GeV: $\psi(3\,^3S_1)$, $\psi(4\,^3S_1)$, $\psi(1\,^3D_1)$, $\psi(2\,^3D_1)$, $\chi_{c2}(2\,^3P_2)$ are accepted in the quark model. The first four are well established states that are used in ref [7] to find the model parameters but in ref. [11] has not used $\psi(4\,^3S_1)$ state in the parameter fitting and included the observed state $\chi_{c2}(2\,^3P_2)$. We compare our results with the experimental decay widths and other theoretical results.

### A. Experimental well established known $c\bar{c}$ states above the 3.73 GeV

#### 1. $\psi(3770)$

The $\psi(3770)$ is generally assigned to be the $\psi(1\,^3D_1)$ $c\bar{c}$ states. It is allowed significantly small $2S$ wave component. It is the first D wave vector charmonium state above the threshold value of 3.73 GeV. The decay mode of $\psi(3770)$ is $D\bar{D}$ and leading as a pure $^3D_1$ state. Its strong decay properties can be explained both with the realistic wave function and simple harmonic oscillator wave function (SHO). The experimental decay width of $\psi(3770)$ is [3]:

$$\Gamma_{\psi(3770)} = 27.2 \pm 1.0 \text{ MeV},$$

and our predicted $^3P_0$ model partial decay width is 28 MeV with SHO wave function. This agree with the experimental decay widths. Its decay widths calculated with realistic wave function is 21 MeV which is close to the experimental average PDG value [3]. BABAR reported [21] the decay width of $\psi(3770)$ as $23.5 \pm 3.7 \pm 0.9$ MeV which is very close to the our calculated value with realistic wave functions.

2. $\psi(4040)$

Next state above the threshold value is $\psi(4040)$. It is exceptionally curiously case to study the strong decay properties. In the quark potential model it is assigned to be $3^3S_1$ charmonium state and its allowed strong decay modes $DD$, $DD^*$, $D^*D^*$ and $D_sD_s$ have been seen in experiments [3]. Its open flavor strong decay widths calculated by us using both realistic and SHO wave functions are reported in table (VIII). The experimental decay width of $\psi(4040)$ is [3]

$$\Gamma_{\psi(4040)} = 80 \pm 10 \text{MeV}.$$

Our calculated decay width of 86 MeV obtained with realistic wave function is in very good agreement with this experimental average PDG value [3] and that calculated with SHO wave function is 66 MeV. This is slightly less than the experimental value [3]. We also compare the partial width ratio of $DD$ and $DD^*$ channels, and of the $D^*D^*$ and $DD^*$ channels calculated using realistic and SHO wave functions with the experimental given ratio. With the realistic wave functions partial width ratio is

$$\frac{\Gamma(DD)}{\Gamma(DD^*)} = 0.1,$$

which is less then the experimental given ratio of $0.24 \pm 0.05 \pm 0.12$ from the BABAR collaboration [22]. The ratio with SHO wave functions is

$$\frac{\Gamma(DD)}{\Gamma(DD^*)} = 0.04.$$

This is not in agreement with BABAR collaboration. Our calculation for realistic wave functions give

$$\frac{\Gamma(D^*D^*)}{\Gamma(DD^*)} = 2.47,$$

and with SHO wave functions we get

$$\frac{\Gamma(D^*D^*)}{\Gamma(DD^*)} = 1.87.$$

These seems to be large as compared to the value $0.18 \pm 0.14 \pm 0.03$ reported by the BABAR collaboration [22]. Other theoretical results [7, 11] have also found values of this ratio as large compared with the measured value from BABAR collaboration.

3. $\psi(4160)$

The $\psi(4160)$ was determined to be $J^{PC} = 1^{--}$. our mass prediction for it is reported in table (III). We have assigned it to be $2^3D_1$ charmonium state in the quark potential model. Its four open charm strong decay channels have been seen in experiment are $DD$, $DD^*$, $D^*D^*$, and $D_sD_s^*$ [3]. Predicted strong decay widths by using the $^3P_0$ model, with both SHO and realistic wave functions are reported in table (VIII). The measured decay width of $\psi(4160)$ is

$$\Gamma_{\psi(4160)} = 70 \pm 10 \text{MeV}.$$

Our theoretical decay widths are large as compared to average measured decay width. But our results agree with the measured decay widths of $107 \pm 8$ MeV for this state from the Crystal Ball (CB) and Bess Data [23]. Its our decay rate of $D_sD_s$ calculated with realistic wave function is nearly small and branching fraction is about 0.06%. This can be the reason why $D_sD_s$ has not seen in experiment and our result agree with ref [11]. Our calculation for the decay channel of $\psi(4160)$ gives

$$\psi(4160) \to D^*D^* > DD > DD^*,$$

which is consistent with other theoretical results [7, 11].

*4. $\psi(4415)$*

$\psi(4415)$ has $J^{PC} = 1^{--}$ and potential model prediction suggests an assignment of $4^3S_1$ charmonium state. Seven open charm strong decays modes are allowed with $c\bar{n}$ for this state ($n = u, d$) and three with $c\bar{s}$ states. Our theoretical results of strong decay width of $\psi(4415)$ are listed in table (VIII). The measured decay width of $\psi(4415)$ is

$$\Gamma_{\psi(4415)} = 70 \pm 20 \text{ MeV}.$$

Our predicted strong decay widths are quit large as compared to the measured value and with the result reported in ref [7, 11]. This difference is mainly due to the meson wave functions. By using the SHO wave function we find the difference to be small. Still results do not agree with the experimental value. But measured value reported in ref [23] $119 \pm 15$ MeV is very close to our predicted value 115 MeV with SHO wave functions.

Using the realistic wave functions, decay rate of $\psi(4415) \to D_s D_s$ is very small and agrees with ref [11]. A Partial decay width ratio with realistic wave function is

$$\frac{\Gamma(DD)}{\Gamma(D^*D^*)} = 0.30,$$

which agrees to the value of 0.29 measured by the BABAR collaboration [22] and agrees with ref [11]. That calculated with SHO wave function is

$$\frac{\Gamma(DD)}{\Gamma(D^*D^*)} = 0.20.$$

This is close to the measured value by the BABAR collaboration [22]. The partial width ratio of $DD^*$ and $D^*D^*$ channel with realistic wave functions is

$$\frac{\Gamma(D^*D)}{\Gamma(D^*D^*)} = 0.1.$$

And with the SHO wave functions, it is

$$\frac{\Gamma(D^*D)}{\Gamma(D^*D^*)} = 0.04,$$

which agrees with the measured value of $0.17 \pm 0.28$ [22] within the range of the data. The largest predicted branching ratio of $\psi(4415)$ using the realistic and SHO wave function is for the $DD_1'$ decay mode, where $D_1'$ is the broader of the two $1^+$ axial mesons near 2.420-2.427 GeV and our result agrees with ref [11]. Finally, difference in calculated decay width with realistic and SHO wave function is mainly due to different mesonic wave functions. That is why our predicted strong decay properties of $\psi(4415)$ is inconsistent with the observed decay width of $\psi(4415)$.

*5. $\chi_{c2}(3930)$*

The first observation of $X(3927)$ was in the $\gamma\gamma \to D\overline{D}$ process by the Belle [24] and BABAR [25] collaboration. This is a good candidate of $\chi_{c2}(2P)$ and quark model predicts an assignment as $2^3P_2$ $c\bar{c}$ state. Its kinematically allowed open charm strong decay modes are $DD$ and $DD^*$. We calculate its strong decay widths both with realistic and SHO wave functions and these are reported in table (VIII). Its observed decay width has been recently reported [3] as

$$\Gamma_{\chi_{c2}(3930)} = 35.3 \pm 2.8 \text{MeV}.$$

Our calculated results with realistic and SHO wave functions do not agree. Its strong decays width using the $^3P_0$ model is 34 MeV with realistic wave function. This is in good agreement with experimental PDG value [3]. The dominant decay mode of $\chi_{c2}(2P)$ is $DD$, while $D^*D$ decay channel is also sizable. Partial width ratio between $D^*D$ and $DD$ calculated with realistic wave function is

$$\frac{\Gamma(D^*D)}{\Gamma(DD)} = 0.54.$$

This agrees with ref. [11]. And with SHO wave functions, it is

$$\frac{\Gamma(D^*D)}{\Gamma(DD)} = 0.66.$$

This is slightly larger than the value with realistic wave function and of ref. [11].

### B. 3S and 4S states

The two unknown open charm pseudoscalar states $3S$ and $4S$ are $3^1S_0$ and $4^1S_0$. Their predicted strong decay widths using the realistic wave functions are 151 MeV and 166 MeV, respectively and with the SHO wave functions are 136 MeV and 113 MeV, respectively. The $X(3940)$ is likely to be $\eta_c(3S)$ with $J^{PC} = 0^{-+}$ [26]. The first observation; $X(3940)$ in $e^-e^+ \to J/\psi X$ is reported by the Belle Collaboration [27] and later on, this states is recognized in the $D^*\overline{D}^*$ invariant mass distribution in the process $e^-e^+ \to J/\psi + D^*\overline{D}^*$ [28]. The updated mass of $X(3940)$ is $M = 3.942^{+7}_{-6} \pm 6$ GeV and its decay width $\Gamma = 37^{+26}_{-15} \pm 8$ MeV. Our predicted mass of $\eta_c(3S)$ is 4.0614 GeV, which is higher than the observed mass and calculated strong decay widths by using realistic and SHO wave function are not in agreement with measurements. The largest branching fraction of the $\eta_c(4S)$ state by using the realistic wave function with S+P combination $\to DD_{2*}$ is about 54%. Ref. [7] has also reported the largest branching ratio for the same decay mode.

### C. 5S state

In the quark model, the mass of $5S$ state is predicted to be in the range of $4.7082 - 4.7284$ GeV. Our predicted mass of $\psi(5S)$ is 4.7284 GeV, which is comparable to the measured value of Y(4660) state of $J^{PC} = 1^{--}$ and we assign it to be $5^3S_1$ state. The Y(4660) was observed in the process $e^-e^+ \to \pi^+\pi^-\psi(2S)$ reported by the Belle collaboration [29]. We study its strong decay width by using the $^3P_0$ decay model and these are reported in table IX. Our calculated partial widths with realistic and SHO wave functions do not agrees. These are listed in table IX. Our predicted decay width is not consistent with the updated measured decay value $\Gamma = 64 \pm 9$ MeV [3]. Ref. [30] reported the decay width of this state as $48 \pm 15 \pm 3$ MeV which is close to the our calculated result 40 MeV with SHO wave functions. And Ref. [31] reported the decay width as $104 \pm 48 \pm 10$ MeV, which is close to the our calculated result with realistic wave functions.

### D. 2P states

#### 1. $\chi_{c1}(2P)$

This state is also known as $X(3872)$ state and was first discovered as reported by Belle in $B$ decays [32], $B \to K\pi^+\pi^-J/\psi$ and was confirmed by BABAR collaboration [33] and first observation in the $J/\psi\pi^+\pi^-$ final state. Its observed mass is reported to be $3871.69 \pm 0.17$ MeV and decay width is $< 1.2$ MeV [3]. This state is the best candidate for $DD^*$ molecule as its mass is close to the threshold mass of $DD^*$ [11]. Many frameworks related to this state had been discussed in recent literature [34–38]. We find that the mass of $\chi_{c1}(2P)$ with our quark model parameters is 3951 MeV which is in agreement with ref. [4] result. We calculate its strong decays width with both realistic and SHO wave functions. These are reported in Table (X). our results with both wave functions are close and are comparable with the other theoretical result [11].

#### 2. $\chi_{c0}(2P)$

An observation of the state $X(3860)$ is reported by Belle Collaboration in the process $e^-e^+ \to J/\psi + DD$ [39] with measured mass

$$M_{\text{exp}} = 3862^{+26+40}_{-32-13} \text{ MeV},$$

and decay width

$$\Gamma_{\text{exp}} = 201^{+154+88}_{-67-82} \text{ MeV}.$$

Our predicted mass of this state is 3.9151 GeV, which is close to the $X(3915)$ state but our result is larger than observed value of $X(3860)$ mass. A charmonium like state X(3915) can be a good candidate for $\chi_{c0}(2P)$ in the quark model. Ref. [40] has also assigned the X(3915) state as a $\chi_{c0}(2P)$. Ref. [41, 42] reported that the $X(3860)$ can be a good candidate for $\chi_{c0}(2P)$. We calculated its the strong decays width in the $^3P_0$ model by using the realistic and SHO wave functions. These are listed in table (X). Our predicted decay width by using the realistic and SHO wave functions are similar i.e. $\Gamma = 2.5 - 3.3$ MeV. Our result is narrow than other theoretical results [7, 11]. This is because of the meson wave functions obtained by using the relativistic effects are incorporated in the potential model. Ref. [11] used the numerical wave function of charmunium state by using the linear potential (LP) model and ref. [7] used the SHO wave function with universal SHO parameter $\beta$. Ref. [41] calculate the $^3P_0$ decay width of the same state and finds a large value of about 110-180 MeV. This shows that more experimental and theoretical studied are required for the further precise understanding of this state.

3. $Z_c(3900)$

The observation of $Z_c(3900)$ is reported by BESIII Collaboration in the process $e^-e^+ \to J/\psi\pi^+\pi^-$ [43]. The updated mass of this state is

$$M_{\text{exp}} = 3888.4 \pm 2.5 \text{ MeV},$$

and the decay width is

$$\Gamma_{\text{exp}} = 28.3 \pm 2.5 \text{ MeV}.$$

Quark model predict the mass of this state as 3938.1 MeV and our $^3P_0$ strong decay widths with realistic and SHO wave function are 68 MeV and 56 MeV, respectively. Our calculated result is larger than the above measured value. Ref. [44] reports the decay width as $63 \pm 24 \pm 26$ MeV, which is a very good agreement to the our predicted result. Our assignment of this state is reported in table (III) on the basis of mass and decay width. Still, understanding of properties of this state, and requires more study for the further identification.

### E. 3P states

1. $\chi_{c2}(3P)$

In the quark model, we assigned the charmonium like state $X(4350)$ as $\chi_{c2}(3P)$. The first observation of this state is reported by Belle in the $\gamma\gamma \to \phi + J/\psi$ mass spectrum with $J^{PC} = 2^{++}$ [45]. Its updated mass and total width are 4350.6 MeV and $13^{+18}_{-9} \pm 4$ MeV, respectively. Our predicted mass of this state is 4315.4 MeV which is close to the observed value. That is why $X(4350)$ is assigned to be a $\chi_{c2}(3P)$ state. Strong decay widths of this state are listed in table (X). Our calculated strong decay with realistic wave function is 10 MeV, which is in good agreement with measured decay width.

2. $\chi_{c1}(3P)$

In the quark model calculation, the predicted mass of $\chi_{c1}(3P)$ is 4.3077 GeV and it is assigned to be $3^3P_1$ state. Its updated measured mass and decay width is 4.274 GeV and $49 \pm 12$ MeV, respectively [3]. We calculated the strong decays width of this state. These are listed in table (X) with both realistic and SHO wave functions. Strong decay width of this state by using realistic and SHO wave fuctions are 28 MeV and 16 MeV, respectively. Our calculated result with realistic wave functions is close to the lower limit of measured value and agrees with other theoretical results [7, 11, 42]. The different partial decay width ratios of the $D^*D$, $D^*D^*$ and $D_sD_s$ channels with realistic wave functions are predicted to be

$$\frac{\Gamma(D^*D^*)}{\Gamma(DD^*)} = 1.3$$

$$\frac{\Gamma(D_sD_s^*)}{\Gamma(DD^*)} = 1.45,$$

and with SHO wave functions as

$$\frac{\Gamma(D^*D^*)}{\Gamma(DD^*)} = 0.25$$

$$\frac{\Gamma(D_s D_s^*)}{\Gamma(DD^*)} = 0.12.$$

Our predicted decay width ratios with realistic and SHO wave functions differ. And our results are also different to other theoretical results [7, 11].

3. $\chi_{c0}(3P)$

$\chi_{c0}(3P)$ state of $J^{PC} = 0^{++}$, has not been established. Our calculation of quark model found the mass of $\chi_{c0}(3P)$ as about 4.2827 GeV, which is larger than observed value. The $^3P_0$ strong decay widths of this states are reported in table (X). Our predicted results with realistic and SHO wave functions are closer to the results of refs. [7, 11].

### F. 1D states

1. $\psi_3(1D)$

In 2003, charmonium like state $X(3842)$ was reported by Belle collaboration [47] and updated measured mass and decay width of this state are $3842.71 \pm 0.16 \pm 0.12$ MeV and $2.79 \pm 0.51 \pm 0.35$ MeV, respectively [3]. In our quark model calculation, $\psi_3(1D)$ mass is predicted as 3.8415 GeV and it is assigned to be $1^3D_3$ state mention in table (III). The strong decays widths of $\psi_3(1D)$ state are reported in table (XI). Our calculated results with SHO wave function and realistic wave functions do not agree. Ref. [11] calculates the decay width of this state to be 0.88 MeV, which agrees to our calculated result of 0.98 MeV with SHO wave functions. Ref. [48] reported the mass of this state to be $3.844^{+0.0675}_{-0.0823} \pm 0.020$ GeV and its decay width ranges from $\Gamma = 5.6 - 6.9$ MeV. This is close to our theoretical results with realistic wave functions.

### G. 2D states

1. $\psi_3(2D)$

$\psi_3(2D)$ state is assigned to be $2^3D_3$ in the quark model prediction. Our strong decay widths of this state are listed in table (XI). The dominant decay channel of this states are $DD$, $DD^*$, $D^*D^*$ and charm strange final meson decay modes $D_s D_s$, $D_s^* D_s^*$ have smaller contribution. Similar result are calculated in ref. [7, 11]. Our calculated results and other theoretical results of the strong decays width have large discrepancy. This suggest that more study is required for the further investigation of this state.

2. $\psi_2(2D)$

In the potential model calculation, $\psi_2(2D)$ mass is predicted to be 4.1764 GeV and it is assigend to be $2^3D_2$ state reported in table (III). GI predicted mass of the same state as 4.208 GeV which is higher than our calculated mass. Our predicted strong decay widths ranges from $\Gamma = 63 - 92$ MeV, which are consistent with others theoretical results [7, 11]. Its dominant decay mode is $D^*D^*$ having a branching ratio of about 59% in our calculations. This is similar for both realist and SHO wave functions.

3. $\eta_{c2}(2D)$

The $2^1D_2$ is a candidate of $\eta_{c2}(2D)$. Our calculated mass of this state is 4.1772 GeV and our strong decays widths of this stats are listed in table (XI). The range of our calculated strong decay widths with both realistic and SHO wave functions are same as the available theoretical results [7, 11]. In this state the dominant decay mode is $D^*D^*$, similar to that of $\psi_2(2D)$, with a branching fraction of about $42\% - 54\%$.

### H. 1F states

There is no experimentally observed state with $1F$ quantum numbers. So, we compare our results with other available theoretical results. There are four multiplets of 1F states $1^3F_4$, $1^3F_3$, $1^3F_2$ and $1^1F_3$. Predicted masses of these states are 4.0888 GeV, 4.0516 Gev, 4.0008 GeV, 4.0561 GeV, respectively. It is found that there are large discrepancies between our predicted masses and the GI model results [4]. We have calculated the strong decay widths of $1F$ states. These are listed in table (XII).

### I. Comparison to other works

In this work, we employ a different model to previous works and conclusion. We have adopted the quark potential model to describe the charmonium, $D$ and $D_s$ mesons. Charmonium sector with relativistic potential model has been studied since long ago [4]. Many systematic studies have been carried out by using the non-relativistic potential [7, 11]. In $c\bar{c}$ strong decays the final state mesons are $D$ and $D_s$. We cannot use the non-relativistic potential model due to the light quark in the final state meson. In that case, we need to approximate meson functions. But in this paper, we have used the potential model by incorporating the relativistic effects for the initial and final meson wave functions to avoid this approximation. Ref. [11] has the recent studies to calculate the strong decays of charmnonium. They have used the linear potential model to find the meson function of $c\bar{c}$ state and approximate the final state meson, using the same value of SHO parameter $(\beta)$ to calculate the strong decay widths. Ref. [7] studied the strong decays of meson by using the same value of the SHO parameter $(\beta)$ for both inital and final mesons states. In comparison, we have calculated the strong decays width by using two different choices of meson function 1) We have used the realistic wave functions for initials and final mesons. 2) We took the different values of $\beta$ for different mesons states. Details of this technique are explained in ref [49]. The comparison of realistic and SHO wave function of $c\bar{c}$ for $S$-wave upto $n = 1, 2, 3, 4, 5$ in position space shown in fig 2. Here the overlap of the realistic and SHO wave functions are almost agree with $n = 1, 2, 3$ but for higher radial excited state the visibility of nodes points are different in radial wave functions. As an example, we plot the $S$-wave $\eta_c$ meson state with realistic and fitted SHO wave functions shown in fig. 2 and overlaps are reported in table VII.

## V. CONCLUSION

In this work, we have first calculated the mass spectrum of charmonium states using the quark potential model by incorporating the relativistic effects. We solved the relativistic type Schrödinger wave equation to find the wave functions of quark anti-quark states. Then, we carried out detail study of strong decays widths of charmonium states using the $^3P_0$ decay model. Some important results found through this study are given as:

1. The mass spectrum of $c\bar{c}$ can be described by quark potential model. Recently assigned states of known $J^{PC}$ are in good agreement with our predicted results.

2. The five well established charmonium states are described in the quark model and large discrepancies found in the strong decay widths by using SHO and realistic wave functions.

3. Strong decay widths of charmonium state $\psi(4S)$ have large differences with realistic and SHO wave functions due to the two different choices of mesons wave functions.

4. In the quark model charmonium like state $X(3940)$ with $J^{PC} = 0^{-+}$ is likely to be $\eta_c(3S)$ state. Its predicted mass and decay width is larger then measured values.

5. The charmonium like state Y(4660) is assigned to be $\psi(5S)$ state on the basis of comparison of our predicted mass and strong decay width with experimental results.

6. The charmonium state $X(3915)$ is a favour of $\chi_0(2P)$. Our predicted mass of this state is close to the measured value of $X(3915)$.

7. In the quark model, $Z_c(3900)$ can be a candidate of $h_c(2p)$ state, although its decay width and mass are larger than the observed value. Our predicted result is in good agreement by using the realistic wave function with the measured value reported in ref [44].

8. The Chamonium like state X(4350) is a candidate of $\chi_2(3P)$ state. Our predicted mass and decays width using realistic wave functions agree with measured values.

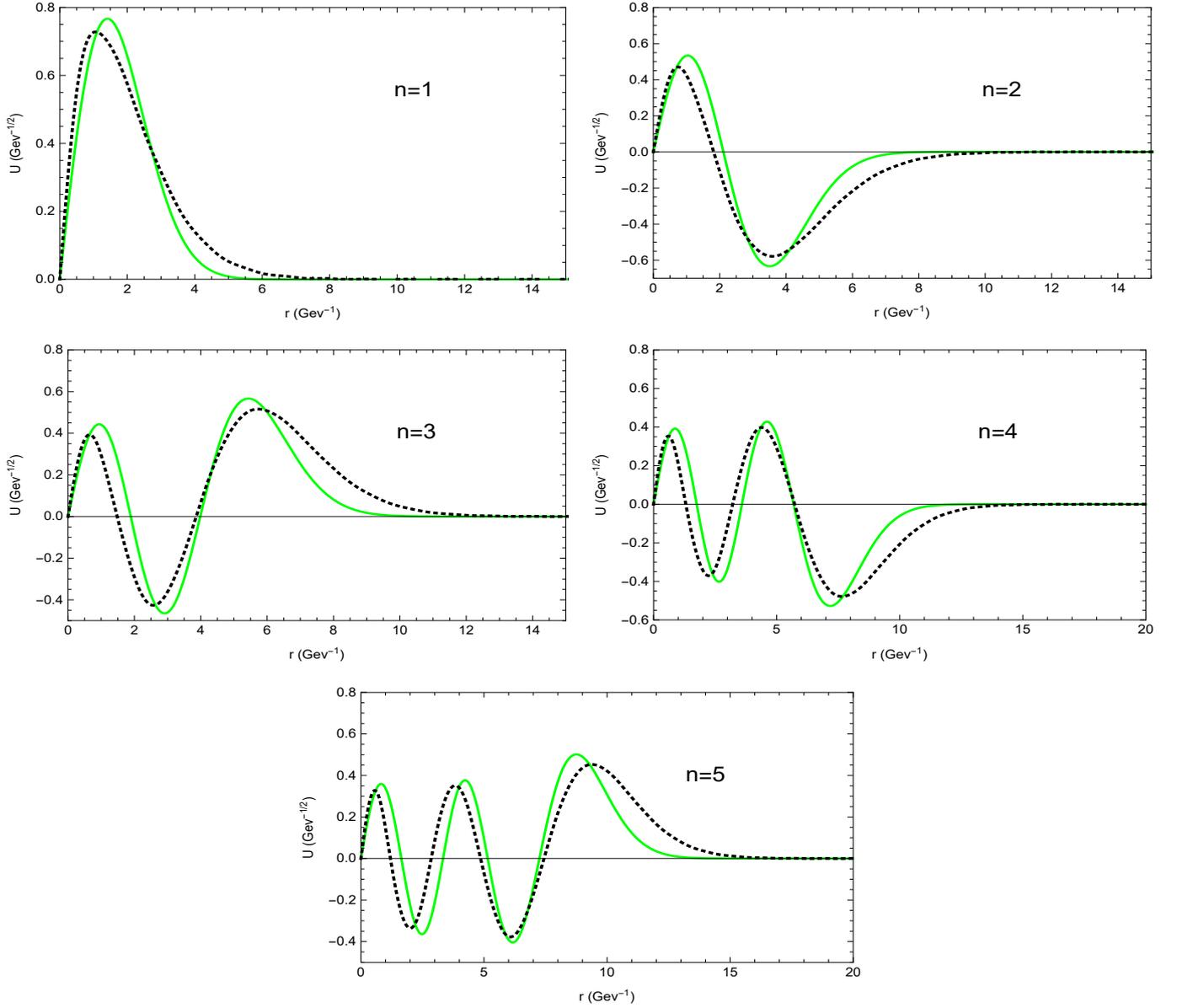

FIG. 2. Radial wave functions of $\eta_c$ states with $n = 1, 2, 3, 4, 5$. Dotted: Realistic wave functions and Solid: Fitted SHO wave functions

9. It is found that, strong decay widths of $\chi_1(3P)$ by using the realistic wave functions are close to the lower limit of measured decay value.

10. The newly charmonium state $X(3842)$ is assigned to be $\psi_3(1D)$ $c\bar{c}$ meson state on the basis of our predicted mass and decay width.

This brief study of spectrum and strong decays width should be helpful to better understand the insight of the charmonium states.

TABLE III. Experimental and theoretical spectrum of $c\bar{c}$ states and SHO fitted $\beta$ values.

| $J^{PC}$ | State | Our calculated mass (GeV) | Expt. (GeV)[3] | NR (GeV) [7] | GI (GeV)[4] | $\beta$ (GeV) |
|---|---|---|---|---|---|---|
| $1^{--}$ | $J/\psi(1^3S_1)$ | 3.1040 | $3.0969 \pm 0.006$ | 3.090 | 3.098 | 0.610 |
| $0^{-+}$ | $\eta_c(1^1S_0)$ | 2.9921 | $2.9839 \pm 0.5$ | 2.982 | 2.975 | 0.708 |
| $1^{--}$ | $\psi'(2^3S_1)$ | 3.6913 | $3.6861 \pm 0.06$ | 3.672 | 3.676 | 0.535 |
| $0^{-+}$ | $\eta_c'(2^1S_0)$ | 3.6346 | $3.6375 \pm 1.1$ | 3.630 | 3.623 | 0.580 |
| $1^{--}$ | $\psi(3^3S_1)$ | 4.0993 | $4.039 \pm 1$ | 4.072 | 4.100 | 0.484 |
| $0^{-+}$ | $\eta_c(3^1S_0)$ | 4.0614 | | 4.043 | 4.064 | 0.507 |
| $1^{--}$ | $\psi(4^3S_1)$ | 4.4340 | $4.421 \pm 4$ | 4.406 | 4.450 | 0.452 |
| $0^{-+}$ | $\eta_c(4^1S_0)$ | 4.4073 | | 4.384 | 4.425 | 0.466 |
| $1^{--}$ | $\psi(5S)$ | 4.7284 | $4.633 \pm 7$ | | | 0.431 |
| $0^{-+}$ | $\eta_c(5S)$ | 4.7082 | | | | 0.440 |
| $1^{--}$ | $\psi(6S)$ | 4.9998 | | | | 0.413 |
| $0^{-+}$ | $\eta_c(6S)$ | 4.9793 | | | | 0.421 |
| $2^{++}$ | $\chi_2(1^3P_2)$ | 3.5321 | $3.5561 \pm 0.07$ | 3.556 | 3.550 | 0.512 |
| $1^{++}$ | $\chi_1(1^3P_1)$ | 3.5042 | $3.5106 \pm 0.05$ | 3.505 | 3.510 | 0.517 |
| $0^{++}$ | $\chi_0(1^3P_0)$ | 3.4464 | $3.4147 \pm 0.30$ | 3.424 | 3.445 | 0.529 |
| $1^{+-}$ | $h_c(1^1P_1)$ | 3.4993 | $3.5254 \pm 0.11$ | 3.516 | 3.517 | 0.530 |
| $2^{++}$ | $\chi_2(2^3P_2)$ | 3.9653 | $3.9222 \pm 1.0$ | 3.972 | 3.979 | 0.475 |
| $1^{++}$ | $\chi_1(2^3P_1)$ | 3.9511 | $3.8717 \pm 0.17$ | 3.925 | 3.953 | 0.480 |
| $0^{++}$ | $\chi_0(2^3P_0)$ | 3.9151 | $3.9184 \pm 1.9$ | 3.852 | 3.916 | 0.491 |
| $1^{+-}$ | $h_c(2^1P_1)$ | 3.9381 | $3.8884 \pm 2.5$ | 3.934 | 3.956 | 0.491 |
| $2^{++}$ | $\chi_2(3^3P_2)$ | 4.3154 | | 4.317 | 4.337 | 0.445 |
| $1^{++}$ | $\chi_1(3^3P_1)$ | 4.3077 | $4.274^{+8}_{-6}$ | 4.271 | 4.317 | 0.449 |
| $0^{++}$ | $\chi_0(3^3P_0)$ | 4.2827 | | 4.202 | 4.292 | 0.458 |
| $1^{+-}$ | $h_c(3^1P_1)$ | 4.2912 | | 4.279 | 4.318 | 0.457 |
| $3^{--}$ | $\psi_3(1^3D_3)$ | 3.8415 | | 3.806 | 3.849 | 0.459 |
| $2^{--}$ | $\psi_2(1^3D_2)$ | 3.8034 | | 3.800 | 3.838 | 0.472 |
| $1^{--}$ | $\psi(1^3D_1)$ | 3.7514 | $3.7737 \pm 0.4$ | 3.785 | 3.819 | 0.485 |
| $2^{-+}$ | $\eta_{c2}(1^1D_2)$ | 3.8096 | | 3.799 | 3.837 | 0.471 |
| $3^{--}$ | $\psi_3(2^3D_3)$ | 4.1976 | | 4.167 | 4.217 | 0.439 |
| $2^{--}$ | $\psi_2(2^3D_2)$ | 4.1764 | | 4.158 | 4.208 | 0.450 |
| $1^{--}$ | $\psi(2^3D_1)$ | 4.1500 | $4.191 \pm 5$ | 4.142 | 4.194 | 0.457 |
| $2^{-+}$ | $\eta_{c2}(2^1D_2)$ | 4.1772 | | 4.158 | 4.208 | 0.449 |
| $3^{--}$ | $\psi_3(3^3D_3)$ | 4.5085 | | $\cdots$ | $\cdots$ | 0.420 |
| $2^{--}$ | $\psi_2(3^3D_2)$ | 4.4947 | | $\cdots$ | $\cdots$ | 0.428 |
| $1^{--}$ | $\psi(3^3D_1)$ | 4.4891 | | $\cdots$ | $\cdots$ | 0.428 |
| $2^{-+}$ | $\eta_{c2}(3^1D_2)$ | 4.4920 | | $\cdots$ | $\cdots$ | 0.429 |
| $4^{++}$ | $\chi_4(1^3F_4)$ | 4.0888 | | 4.021 | 4.095 | 0.425 |
| $3^{++}$ | $\chi_3(1^3F_3)$ | 4.0516 | | 4.029 | 4.097 | 0.443 |
| $2^{++}$ | $\chi_2(1^3F_2)$ | 4.0008 | | 4.029 | 4.092 | 0.462 |
| $3^{+-}$ | $h_{c3}(1^1F_3)$ | 4.0561 | | 4.026 | 4.094 | 0.443 |

[a] GI means Godfrey-Isgur potential model [4]

TABLE IV. Experimental and theoretical masses of $c\bar{u}$ used in strong decays of $c\bar{c}$ with SHO fitted $\beta$ values.

| Meson state | | Mass$_{(Exp)}$ (GeV) | Mass$_{(theo)}$ (GeV) | $\beta$ (GeV) |
|---|---|---|---|---|
| $D$ | $1^1S_0$ | 1.86726 | 1.8586 | 0.564 |
| $D^*$ | $1^3S_1$ | 2.0085 | 2.0021 | 0.437 |
| $D_0^*$ | $1^3P_0$ | 2.3245 | 2.2886 | 0.491 |
| $D_1$ | $1P$ | 2.4208 | 2.4028 | 0.400($1^1P_1$),0.404($1^3P_1$) |
| $D_2^*$ | $1^3P_2$ | 2.46305 | 2.4670 | 0.363 |
| $D_1'$ | $1P'$ | 2.4270 | 2.4760 | 0.399($1^1P_1$),0.403($1^3P_1$) |

TABLE V. Experimental and theoretical masses of $c\bar{s}$ used in strong decays of $c\bar{c}$ with SHO fitted $\beta$ values.

| Meson state | | Mass$_{(Exp)}$ (GeV) | Mass$_{(theo)}$(GeV) | $\beta$ (GeV) |
|---|---|---|---|---|
| $D_s$ | $1^1S_0$ | 1.96834 | 1.9720 | 0.594 |
| $D_s^*$ | $1^3S_1$ | 2.1122 | 2.1057 | 0.474 |
| $D_{s0}^*$ | $1^3P_0$ | 2.3178 | 2.3276 | 0.547 |
| $D_{s1}$ | $1^3P_1$ | 2.4595 | 2.5047 | 0.413($1^1P_1$),0.425($1^3P_1$) |
| $D_{s2}^*$ | $1^3P_2$ | 2.5691 | 2.5691 | 0.378 |
| $D_{s1'}$ | $1^1P_1$ | 2.5311 | 2.5484 | 0.418($1^1P_1$),0.432($1^3P_1$) |

| Generic Decay | Subprocess | $I_{flavor}(d_1)$ | $I_{flavor}(d_2)$ | $\mathcal{F}$ |
|---|---|---|---|---|
| $X_c \to D\bar{D}$ | $X_c \to D^+D^-$ | 0 | 1 | 2 |
| $X_c \to D^*\bar{D}$ | $X_c \to D^{*+}D^-$ | 0 | 1 | 4 |
| $X_c \to D^*\bar{D}^*$ | $X_c \to D^{*+}D^{*-}$ | 0 | 1 | 2 |
| $X_c \to D_s\bar{D}_s$ | $X_c \to D_s^+D_s^-$ | 0 | 1 | 1 |
| $X_c \to D_s^*\bar{D}_s$ | $X_c \to D_s^{*+}D_s^-$ | 0 | 1 | 2 |
| $X_c \to D_s^*\bar{D}_s^*$ | $X_c \to D_s^{*+}D_s^{*-}$ | 0 | 1 | 1 |

TABLE VI. Flavor factors for charmonium decay, where $|X_c\rangle \equiv |c\bar{c}\rangle$.

TABLE VII. Overlap $= \int U_{\text{Realistic}} U_{\text{SHO}} dr$ of $\eta_c$ mesons for realistic and SHO wave function.

| $n$ | Overlap (%) |
|---|---|
| 1 | 98.7 |
| 2 | 95.2 |
| 3 | 93 |
| 4 | 89 |
| 5 | 86.7 |

TABLE VIII. Strong decay widths of five known $c\bar{c}$ states above the 3.73 GeV, $\psi(3\ ^3S_1)$, $\psi(4\ ^3S_1)$, $\psi(1\ ^3D_1)$, $\psi(2\ ^3D_1)$, $\chi_{c2}(2\ ^3P_2)$.

| Meson State | Decay Mode | $\Gamma_{exp}$ | $\Gamma_{Realistic}(MeV)$ | Br(%) | $\Gamma_{SHO}(MeV)$ | Br(%) | $\Gamma_{the}(MeV)$[11] | Br(%) | $\Gamma_{the}(MeV)$[7] | Br(%) |
|---|---|---|---|---|---|---|---|---|---|---|
| $\psi(3\ ^3S_1)$ | DD | | 2.2 | 2.6 | 0.88 | 0.12 | 2.1 | 3.6 | 0.1 | 0.14 |
| | DD* | | 24 | 28 | 23 | 35 | 10.7 | 18 | 33 | 45 |
| | D*D* | | 58 | 67 | 42 | 64 | 41 | 69 | 33 | 45 |
| | $D_sD_s$ | | 2.3 | 2.7 | 0.30 | 0.45 | 5.9 | 9.0 | 7.8 | 11 |
| | Total | 80 MeV | 86 | 100 | 66 | 100 | 60 | 100 | 74 | 100 |
| $\psi(4\ ^3S_1)$ | DD | | 6.04 | 2.96 | 3.0 | 2.6 | 1.7 | 2.6 | 0.4 | 0.5 |
| | DD* | | 1.92 | 0.94 | 0.47 | 0.41 | 1.0 | 1.5 | 2.3 | 2.9 |
| | D*D* | | 20 | 10 | 14 | 12 | 3.8 | 5.7 | 16 | 21 |
| | $DD_1$ | | 60 | 29 | 41 | 36 | 12 | 18 | 31 | 40 |
| | $DD_1'$ | | 62 | 30 | 42 | 37 | 16 | 24 | 1.0 | 1.2 |
| | $DD_2^*$ | | 44 | 22 | 14 | 12 | 17 | 25 | 23 | 29 |
| | $D^*D_0^*$ | | 3.5 | 1.7 | 0.00 | 0.0 | 8.7 | 13 | 0.0 | 0.0 |
| | $D_sD_s$ | | 0.04 | 0.02 | 0.30 | 0.26 | 0.009 | 0.1 | 1.3 | 1.6 |
| | $D_sD_s^*$ | | 3.3 | 1.6 | 0.31 | 0.27 | 3.3 | 5.0 | 2.6 | 3.3 |
| | $D_s^*D_s^*$ | | 3.2 | 1.57 | 0.30 | 0.26 | 3.8 | 5.7 | 0.7 | 0.8 |
| | Total | 62 MeV | 204 | 100 | 115 | 100 | 66 | 100 | 61 | 100 |
| $\psi(1\ ^3D_1)$ | DD | 27 MeV | 21 | 100 | 28 | 100 | 27 | 100 | 43 | 100 |
| $\psi(2\ ^3D_1)$ | DD | | 24 | 19 | 32 | 32 | 12 | 15.4 | 16 | 22 |
| | DD* | | 3.3 | 2.6 | 3.2 | 3.2 | 2.6 | 3.4 | 0.4 | 0.67 |
| | D*D* | | 93 | 73 | 59 | 60 | 43 | 69 | 35 | 47 |
| | $D_sD_s$ | | 0.08 | 0.06 | 2.5 | 2.5 | 0.6 | 0.2 | 8.0 | 11 |
| | $D_sD_s^*$ | | 7.2 | 5.6 | 2.6 | 2.6 | 10 | 13 | 14 | 19 |
| | Total | 70 MeV | 128 | 100 | 99 | 100 | 79 | 100 | 74 | 100 |
| $\chi_{c2}(2\ ^3P_2)$ | DD | | 22 | 65 | 9.9 | 60 | 24 | 75 | 42 | 53 |
| | DD* | | 12 | 35 | 6.6 | 40 | 14 | 26 | 37 | 47 |
| | Total | 35 MeV | 34 | 100 | 16 | 100 | 38 | 100 | 80 | 100 |

TABLE IX. Strong decay widths of the 3S, 4S and 5S of $c\bar{c}$ mesons.

| Meson State | Decay Mode | $\Gamma_{Realistic}(MeV)$ | Br(%) | $\Gamma_{SHO}(MeV)$ | Br(%) | $\Gamma_{the}(MeV)$[11] | Br(%) | $\Gamma_{the}(MeV)$[7] | Br(%) |
|---|---|---|---|---|---|---|---|---|---|
| $\eta_c(3\,^1S_0)$ | $DD^*$ | 34 | 23 | 46 | 34 | 21 | 28 | 47 | 59 |
| | $D^*D^*$ | 117 | 77 | 90 | 66 | 54 | 72 | 33 | 41 |
| | Total | 151 | 100 | 136 | 100 | 75 | 100 | 80 | 100 |
| $\eta_c(4\,^1S_0)$ | $DD^*$ | 2.1 | 1.2 | 0.21 | 0.2 | 0.3 | 0.5 | 6.3 | 10.3 |
| | $D^*D^*$ | 28 | 17 | 81 | 72 | 5.6 | 8.5 | 14 | 22.9 |
| | $DD_0^*$ | 25 | 15 | 3.2 | 2.8 | 24 | 36.7 | 11 | 18.0 |
| | $DD_2^*$ | 90 | 54 | 27 | 24 | 28 | 43.1 | 24 | 39.3 |
| | $D_sD_s^*$ | 6.2 | 3.7 | 1.0 | 0.9 | 5.7 | 8.7 | 2.2 | 3.6 |
| | $D_s^*D_s^*$ | 1.2 | 0.7 | 0.0 | 0.0 | 1.6 | 2.5 | 2.2 | 3.6 |
| | $D_sD_{s0}^*$ | 13 | 7.8 | 0.27 | 0.2 | ... | ... | 0.6 | 0.98 |
| | Total | 166 | 100 | 113 | 100 | 66 | 100 | 61 | 100 |
| $\psi(5\,^3S_1)$ | $DD$ | 2.02 | 1.9 | 1.4 | 3.5 | 0.6 | 1.1 | | |
| | $DD^*$ | 6.2 | 5.7 | 0.52 | 1.3 | 1.2 | 2.1 | | |
| | $D^*D^*$ | 14 | 13 | 2.63 | 6.6 | 0.0 | 0.0 | | |
| | $DD_1$ | 19 | 17 | 0.25 | 0.62 | 0.06 | 0.1 | | |
| | $DD_1'$ | 19 | 17 | 0.23 | 0.58 | 10 | 17.5 | | |
| | $DD_2^*$ | 15 | 14 | 1.76 | 4.4 | 0.12 | 0.2 | | |
| | $D^*D_0^*$ | 5.2 | 4.8 | 0.77 | 1.9 | 11 | 18.3 | | |
| | $D^*D_1$ | 9.1 | 8.3 | 1.26 | 3.1 | 4.3 | 7.3 | | |
| | $D^*D_1'$ | 0.1 | 0.09 | 1.32 | 3.3 | 18 | 30.3 | | |
| | $D^*D_2^*$ | 0.88 | 1.7 | 12 | 30 | 12 | 20.1 | | |
| | $D_sD_s$ | 0.0 | 0.0 | 0.27 | 0.68 | 0.0 | 0.0 | | |
| | $D_sD_s^*$ | 0.95 | 0.87 | 0.0 | 0.0 | 0.3 | 0.5 | | |
| | $D_s^*D_s^*$ | 1.03 | 0.94 | 0.12 | 0.30 | 1.5 | 2.5 | | |
| | $D_sD_{s1}$ | 0.9 | 0.83 | 0.32 | 0.80 | ... | ... | | |
| | $D_sD_{s1}'$ | 1.1 | 1.01 | 0.23 | 0.58 | ... | ... | | |
| | $D_sD_{s2}^*$ | 2.3 | 2.1 | 0.46 | 1.2 | ... | ... | | |
| | $D_s^*D_{s0}^*$ | 0.6 | 0.6 | 0.0 | 0.0 | ... | ... | | |
| | $D_s^*D_{s1}$ | 11 | 10 | 16 | 40 | ... | ... | | |
| | Total | 109 | 100 | 40 | 100 | 58 | 100 | | |
| $\eta_c(5\,^1S_0)$ | $DD^*$ | 7.8 | 2.6 | 1.5 | 2.2 | 0.5 | 0.7 | | |
| | $D^*D^*$ | 5.4 | 1.8 | 2.04 | 3.0 | 0.4 | 0.6 | | |
| | $DD_0^*$ | 14 | 4.6 | 0.21 | 0.31 | 13 | 18.9 | | |
| | $DD_2^*$ | 28 | 9.3 | 14.8 | 22 | 2.0 | 3.0 | | |
| | $D^*D_1$ | 101 | 33 | 25 | 37 | 11 | 15.9 | | |
| | $D^*D_1'$ | 97 | 32 | 21 | 31 | 26 | 41.3 | | |
| | $D^*D_2^*$ | 35.6 | 12 | 2.4 | 3.5 | 11 | 16.1 | | |
| | $D_sD_s^*$ | 1.3 | 0.43 | 0.001 | 0.001 | 0.8 | 1.3 | | |
| | $D_s^*D_s^*$ | 3.1 | 1.0 | 0.3 | 0.44 | 1.4 | 2.1 | | |
| | $D_sD_{s0}^*$ | 9.1 | 3.0 | 0.16 | 0.24 | – | – | | |
| | Total | 302 | 100 | 68 | 100 | 67 | 100 | | |

TABLE X. Strong decays widths of the 2P and 3P states of $c\bar{c}$ mesons.

| Meson State | Decay Mode | $\Gamma_{Realistic}(MeV)$ | $Br(\%)$ | $\Gamma_{SHO}(MeV)$ | $Br(\%)$ | $\Gamma_{the}(MeV)$[11] | $Br(\%)$ | $\Gamma_{the}(MeV)$[7] | $Br(\%)$ |
|---|---|---|---|---|---|---|---|---|---|
| $\chi_1(2\,^3P_1)$ | $DD^*$ | 86.11 | 100 | 74.52 | 100 | 102 | 100 | 165 | 100 |
| $\chi_0(2\,^3P_0)$ | $DD$ | 2.52 | 100 | 2.37 | 100 | 22 | 100 | 30 | 100 |
| $h_c(2\,^1P_1)$ | $DD^*$ | 68 | 100 | 56 | 100 | 64 | 100 | 87 | 100 |
| $\chi_2(3\,^3P_2)$ | $DD$ | 0.14 | 1.4 | 3.13 | 11 | 8.1 | 19 | 8.0 | 12 |
| | $DD^*$ | 3.1 | 31 | 7.3 | 26 | 17 | 40 | 2.4 | 3.6 |
| | $D^*D^*$ | 5.9 | 59 | 13 | 46 | 4.2 | 9.8 | 24 | 36 |
| | $DD_1$ | 0.0 | 0.0 | 0.94 | 3.4 | 1.0 | 2.3 | 1.1 | 1.7 |
| | $DD_1'$ | 0.0 | 0.0 | 0.73 | 2.6 | 7.3 | 17 | 12 | 18 |
| | $D_sD_s$ | 0.15 | 1.5 | 0.42 | 1.5 | 1.0 | 2.0 | 0.8 | 1.2 |
| | $D_sD_s^*$ | 0.22 | 2.2 | 0.09 | 0.32 | 0.3 | 0.7 | 11 | 17 |
| | $D_s^*D_s^*$ | 0.85 | 8.5 | 2.46 | 8.8 | 4.8 | 11 | 7.2 | 11 |
| | Total | 10 | 100 | 28 | 100 | 43 | 100 | 66 | 100 |
| $\chi_1(3\,^3P_1)$ | $DD^*$ | 6.1 | 22 | 7.6 | 48 | 7.1 | 31 | 6.8 | 17 |
| | $D^*D^*$ | 8.1 | 29 | 1.9 | 12 | 0.2 | 0.8 | 19 | 49 |
| | $DD_0^*$ | 0.0 | 0.0 | 0.01 | 0.1 | 0.001 | 0.004 | 0.1 | 0.2 |
| | $D_sD_s^*$ | 8.9 | 32 | 0.9 | 5.6 | 11 | 48 | 9.7 | 9.7 |
| | $D_s^*D_s^*$ | 4.8 | 17 | 5.8 | 36 | 5.5 | 24 | 2.7 | 2.7 |
| | Total | 28 | 100 | 16 | 100 | 23 | 100 | 39 | 100 |
| $\chi_0(3\,^3P_0)$ | $DD$ | 16 | 33 | 17 | 35 | 0.04 | 0.1 | 0.5 | 1.0 |
| | $D^*D^*$ | 23 | 47 | 26 | 53 | 21 | 64 | 43 | 84 |
| | $D_sD_s$ | 0.26 | 0.5 | 1.1 | 2.2 | 8.9 | 27 | 6.8 | 13 |
| | $D_s^*D_s^*$ | 9.8 | 20 | 4.6 | 9.4 | 2.7 | 8.0 | ... | |
| | Total | 49 | 100 | 49 | 100 | 33 | 100 | 51 | 100 |
| $h_c(3\,^1P_1)$ | $DD^*$ | 16 | 28 | 3.5 | 11 | 14 | 32 | 3.0 | 4.0 |
| | $D^*D^*$ | 19 | 33 | 18 | 58 | 4.8 | 11 | 22 | 29 |
| | $DD_0^*$ | 9.3 | 16 | 4.8 | 15 | 15 | 34 | 28 | 37 |
| | $D_sD_s^*$ | 7.8 | 13 | 1.5 | 4.8 | 6.5 | 15 | 15 | 20 |
| | $D_s^*D_s^*$ | 5.6 | 10 | 3.2 | 10 | 3.6 | 8.0 | 7.5 | 10 |
| | Total | 58 | 100 | 31 | 100 | 44 | 100 | 75 | 100 |

TABLE XI. Strong decays widths of the 1D and 2D states of $c\bar{c}$ mesons.

| Meson State | Decay Mode | $\Gamma_{Realistic}(MeV)$ | $Br(\%)$ | $\Gamma_{SHO}(MeV)$ | $Br(\%)$ | $\Gamma_{the}(MeV)$[11] | $Br(\%)$ | $\Gamma_{the}(MeV)$[7] | $Br(\%)$ |
|---|---|---|---|---|---|---|---|---|---|
| $\psi_3(1\,^3D_3)$ | $DD$ | 13 | 100 | 0.98 | 100 | 0.88 | 100 | $\cdots$ | $\cdots$ |
| $\psi_3(2\,^3D_3)$ | $DD$ | 6.2 | 6.1 | 0.4 | 0.7 | 7.2 | 11 | 24 | 16 |
|  | $DD^*$ | 46 | 46 | 20 | 36 | 29 | 45 | 50 | 34 |
|  | $D^*D^*$ | 43 | 43 | 34 | 61 | 20 | 31 | 67 | 45 |
|  | $D_sD_s$ | 3.1 | 3.1 | 0.25 | 0.4 | 5.5 | 8.5 | 5.7 | 3.8 |
|  | $D_sD_s^*$ | 3.1 | 3.1 | 1.0 | 1.8 | 2.9 | 4.5 | 1.2 | 0.8 |
|  | Total | 101 | 100 | 56 | 100 | 65 | 100 | 148 | 100 |
| $\psi_2(2\,^3D_2)$ | $DD^*$ | 36 | 39 | 22 | 35 | 28 | 38 | 34 | 37 |
|  | $D^*D^*$ | 54 | 59 | 37 | 59 | 28 | 38 | 32 | 35 |
|  | $D_sD_s^*$ | 2.2 | 2.4 | 4.4 | 7.0 | 18 | 24 | 26 | 28 |
|  | Total | 92 | 100 | 63 | 100 | 74 | 100 | 92 | 100 |
| $\eta_{c2}(2\,^1D_2)$ | $DD^*$ | 53 | 52 | 29 | 41 | 37 | 49 | 50 | 45 |
|  | $D^*D^*$ | 43 | 42 | 38 | 54 | 25 | 33 | 43 | 39 |
|  | $D_sD_s^*$ | 5.6 | 5.5 | 3.2 | 4.6 | 13 | 18 | 18 | 16 |
|  | Total | 102 | 100 | 70 | 100 | 75 | 100 | 111 | 100 |

TABLE XII. Strong decays widths of the 1F states of $c\bar{c}$ mesons.

| Meson State | Decay Mode | $\Gamma_{Realistic}(MeV)$ | $Br(\%)$ | $\Gamma_{SHO}(MeV)$ | $Br(\%)$ | $\Gamma_{the}(MeV)$[7] | $Br(\%)$ |
|---|---|---|---|---|---|---|---|
| $\chi_4(1\,^3F_4)$ | $DD$ | 12 | 24 | 7.1 | 15 | 6.8 | 82 |
| | $DD^*$ | 9.7 | 19 | 6.5 | 14 | 1.4 | 17 |
| | $D^*D^*$ | 28 | 56 | 34 | 71 | 0.05 | 0.60 |
| | $D_sD_s$ | 0.25 | 0.50 | 0.11 | 0.23 | 0.02 | 0.24 |
| | $D_sD_s^*$ | 0.0 | 0.0 | 0.0 | 0.0 | — | |
| | Total | 50 | 100 | 48 | 100 | 8.3 | 100 |
| $\chi_3(1\,^3F_3)$ | $DD^*$ | 58 | 97 | 73 | 97 | 83 | 99 |
| | $D^*D^*$ | 1.8 | 3 | 2.0 | 2.7 | 0.2 | 0.24 |
| | Total | 60 | 100 | 75 | 100 | 84 | 100 |
| $\chi_2(1\,^3F_2)$ | $DD$ | 95 | 82 | 49 | 64 | 98 | 61 |
| | $DD^*$ | 21 | 18 | 26 | 34 | 57 | 42 |
| | $D_sD_s$ | 0.02 | 0.02 | 1.4 | 1.8 | 5.9 | 3.7 |
| | Total | 116 | 100 | 76 | 100 | 161 | 100 |
| $h_{c3}(1\,^1F_3)$ | $DD^*$ | 48 | 92 | 56 | 95 | 61 | 99 |
| | $D^*D^*$ | 3.7 | 7.2 | 3.1 | 5.0 | 0.1 | 0.16 |
| | Total | 52 | 100 | 59 | 100 | 61 | 100 |